\documentclass{emulateapj}

\begin{document}

\submitted{To be submitted to ApJ}
\title{The unusual smoothness of the extragalactic unresolved radio background}

\author{Gilbert P. Holder}
\affil{Department of Physics, McGill University, Montreal QC H3A 2T8}

\begin{abstract}
If the radio background is coming from cosmological sources, there should be some 
amount of clustering due to the large scale structure in the universe. Simple models
for the expected clustering combined with the recent measurement by ARCADE-2 of the
mean extragalactic temperature lead to predicted clustering levels that are substantially 
above upper limits from searches for anisotropy on arcminute scales
using ATCA and the VLA. The rms
temperature variations in the cosmic radio background appear to be more than a factor of 
10 smaller (in temperature) than the fluctuations in the cosmic infrared background.
It is therefore 
extremely unlikely that this background comes from galaxies, galaxy clusters, or
any sources that trace dark matter halos at $z \la 5$, unless typical sources are
smooth on arcminute scales, requiring typical sizes of several Mpc.  
\end{abstract}

\section{Introduction}

Recent observations of the extragalactic radio background \citep{fixsen11} have
provided a precise measurement of the temperature of the extragalactic sky as a function of frequency at GHz frequencies. 
The measured temperature is roughly an order of magnitude larger
than expectations based on extrapolations of faint source counts \citep{gervasi08}.
This relatively high extragalactic radio temperature has proven difficult to understand 
theoretically \citep{seiffert11, singal10}, including attempts to link the radio
background to dark matter annihilations \citep{fornengo11, hooper12}.

Given that the mean temperature is difficult to understand, a natural next question
is whether the variance can be understood, i.e., the clustering of the radio background.
Clustering of the cosmic infrared background (CIB) has been recently measured by
many experiments \citep{viero09, hall10, planck11}, and these measurements have
provided new insights into the sources that contribute to the CIB \citep{penin12}.
Clustering measurements of the cosmic radio background (hereafter, CRB)
can provide similar information about the sources that contribute.

In this short letter we first review the observed upper limits on clustering at
GHz frequencies, then calculate the expected levels of clustering in some simple
scenarios. We will show that
the expected level of clustering (rms) is substantially larger than observed
upper limits, requiring either that the emission comes from high redshifts
($z \ga 5$), or is coming from sources that are extremely smooth on arcminute scales
(typical source sizes of a few Mpc or larger).

\section{Observed Upper Limits on Clustering}

To date, there have been measurements of the angular power spectrum of resolved sources
\citep{blake04}, but there exist only upper limits on fluctuations of the unresolved
radio background. 

Searches for cosmic microwave background (CMB) anisotropies at low frequencies 
provide the strongest constraints on clustering of the radio background. Searches have been
done for anisotropies using the Very Large Array (VLA) at frequencies of
4.86 GHz \citep{fomalont88} and 8.4 GHz \citep{partridge97}, 
and using the Australia Telescope Compact Array (ATCA)
at 8.7 GHz \citep{subrahmanyan00}. 
No anisotropy was detected in any of these searches, yielding only upper limits,
shown in Table 1.

These limits on CMB fluctuations can also be used as limits on CRB clustering.
To convert to fractional variations in the radio background, we simply need to
replace the normalization by $T_{cmb}$ with the extragalactic radio 
background temperature as a function of frequency $\nu$.

The ARCADE-2 experiment recently reported measurements of the radio
background at frequencies of 3,8 and 10 GHz \citep{fixsen11} and combined
these data with measurements in the literature to determine the
extragalactic radio temperature from 22 MHz to 10 GHz:
\begin{equation}
T_{arcade} = 1.26\pm 0.09 {\rm K} \bigl({\nu \over {\rm GHz}} \bigr)^{-2.6\pm 0.04} \ .
\end{equation}

Using extrapolations of known source counts \citep{gervasi08}, the expected temperature
of the radio sky is
\begin{equation}
T_{counts} = 0.23 {\rm K} \bigl({\nu \over {\rm GHz}} \bigr)^{-2.7} \ .
\end{equation}

The residual background from unresolved sources is defined to be
$T_{excess}=T_{arcade}-T_{counts}$. CMB anisotropy searches generally
either remove or avoid bright point sources; for these purposes we assume that
all contributions from known sources have been removed, which will in practice
be a slight overestimate of source removal efficiency. This will in turn lead
to overly conservative upper limits on the clustering of the CRB.
Upper limits on the CRB (using $T_{excess}$) are shown in
Table 1 for the various experiments. The most stringent constraints are 
from the 4.9 GHz VLA experiment and ATCA at 8.7 GHz, requiring rms fluctuations
in the CRB to be below 1\% on arcminute scales.

\begin{table}
\begin{tabular}{|c|c|c|c|c|}
\hline
 &  & \multicolumn{3}{|c|}{$95\%$ confidence upper limits}   \\
Frequency & $\theta('')$ & $dT/T_{cmb}$ & $dT/T_{arcade}$ &  $dT/T_{excess}$ \\
\hline
\hline
4.86 GHz & 12 & $8.5\times 10^{-4}$ & 0.11  & 0.13   \\
Fomalont et al	& 18 & $1.2\times 10^{-4}$ & 0.016 & 0.019  	\\
(1988)	   & 30 & $8\times 10^{-5}$ &  0.011 & 0.013   	\\
	   & 60 & $6\times 10^{-5}$ &  0.008 & 0.009   	\\
\hline
8.4 GHz & 6 & $1.3\times 10^{-4}$ & 0.070 & 0.082   \\
Partridge et al & 10 & $7.9\times 10^{-5}$ & 0.043 & 0.051 \\
  (1997)        & 18 & $4.8\times 10^{-5}$ & 0.026 & 0.031   	\\
          & 30 & $3.5\times 10^{-5}$ & 0.019 & 0.023   	\\
  	  & 60 & $2.0\times 10^{-5}$ & 0.011 & 0.013   	\\
	  & 80 & $2.1\times 10^{-5}$ & 0.011 & 0.014   	\\
\hline
8.7 GHz & 120 & $1.4\times 10^{-5}$ & 0.0084 & 0.0099  \\
Subrahmanyan & & & & \\
et al (2000) & & & & \\
\hline
\end{tabular}
\caption{Upper limits (95\% confidence) on the fractional radio background anisotropy 
obtained from 
CMB anisotropy limits (column 3) using the temperature measured by
the ARCADE-2 experiment (column 4) and corrected for known 
source populations (column 5). } 
\end{table}

As a point of comparison, the recent measurements of clustering of the
cosmic infrared background (CIB) using the Planck experiment find that 
$\Delta T/T_{CIB} \sim 0.1-0.15$ at mm-wave frequencies \citep{planck11}. 
Whatever is contributing to the extragalactic radio background must  
have rms fluctuations that are more than an order of magnitude smaller than the sources
that contribute to the CIB. 

\section{Expected Clustering of the Radio Background}

Calculating the expected clustering of the CRB is difficult to do at
high precision, since the redshift distribution of the sources
is not well known and it is not known how radio sources trace dark matter halos.
However, it is possible to roughly estimate the expected amount of clustering
for different assumptions about the typical masses and redshifts for the radio sources
that could contribute to the radio background.

We adopt a simple linear bias model \citep{kaiser84}, 
where it is assumed that fluctuations in the
number density of galaxies are proportional to the fluctuations in the matter density
$\Delta n/n = b \Delta \rho/\rho$. The constant of proportionality depends on the
mass and redshift of the dark matter halos that host the sources of interest. For rare
objects, one expects $b$ to be substantially greater than 1, while it is possible for
$b$ to be less than 1 for very low mass objects. However, it is not expected that $b\la 0.5$
for any population that traces dark matter halos \citep{mo96, seljak04}. 
For simplicity, in what follows we assume $b=1$ at all redshifts for the sources that 
make up the radio background.

To calculate the expected CRB fluctuations, we follow the procedure for calculating 
CIB fluctuations laid out in \citet{haiman00}.
The distribution in redshifts that contribute to the extragalactic radio background is
not well known. If we write the fraction of the radio background contributed by a 
distance $d\chi$ in comoving distance as $df/d\chi$, then the angular power spectrum
of temperature fluctuations as a function of multipole number $\ell$ can be obtained using the Limber approximation \citep{kaiser92} as
\begin{equation}
C_{\ell}(\ell) = \int d\chi {1 \over \chi^2} \bigl( {df \over d\chi} \bigr)^2
	P({\ell \over \chi}, \chi) \ ,
\end{equation}
where $P(k, \chi)$ is the matter power spectrum at wavenumber $k$ at comoving distance
$\chi$. For these calculations we use the linear power spectrum, ignoring the increased
power expected in the presence of non-linear evolution. On the scales of interest,
non-linearity could increase the predicted power by large amounts;  for this
work we are most interested in lower limits to the predicted power, given the 
remarkable apparent smoothness of the radio sky.

In the absence of a specific model for the radio background, we investigate the
simplest $df/d\chi$: a top hat function in comoving distance, where it is assumed that
the cosmic radio background is generated uniformly in comoving distance between some
redshifts $z_{min}$ and $z_{max}$. Predicted angular power spectra 
are shown in Figure 1, along with the
limits from Table 1. To convert angle to multipole for the observed upper limits, 
we use $\ell \sim 2.35/\theta$, where $\theta$ is the FWHM of the beam in radians.
The observed upper limits are in clear tension with theoretical expectations for
a normal population of galaxies.

As a cross-check on this simple model, we can use observed clustering. For the CIB,
the fractional fluctuations found by Planck at 845 GHz are of order 10-15\%
\citep{planck11}, consistent with a population of galaxies with bias factors around
two and/or a small amount of non-linear evolution. Using resolved galaxies in the
NVSS survey, the fluctuations in radio galaxy number density
on degree scales were found to be of order
3.5\% at $l\sim 100$ \citep{blake04}, consistent with the predictions in  
Figure 1, especially within the large uncertainties of redshift distributions
and unknown bias factors. 

To get extremely low clustering amplitudes, it is required to have a broad range in
redshifts contributing (to have more averaging of fluctuations along the line of sight) 
and/or to have contributions from higher $z$ (where the matter power
spectrum amplitude is lower due to the growth of structure).

Another way to suppress the small-scale clustering is to have the sources be 
intrinsically large, with very little structure on the arcminute scales
probed by the ATCA and VLA anisotropy searches.
To model this, we assume that each source samples the underlying dark matter density,
but has a spatial profile described by a Gaussian radial profile, which leads to
a smoothing of the matter power spectrum $P(k)_{smooth}=P(k)\exp(-k^2 \sigma^2)$,
where $\sigma={\rm FWHM_{smooth}}/2.35$. 
For the CRB sources to be smooth enough on small scales
to have rms fluctuations smaller than the ATCA limits
requires the sources to be larger than $\sim 2 h^{-1}$Mpc, as shown by the dotted
lines in Figure 1. 

Shocks from the formation of large scale structure have been suggested as a source of
fluctuations in the radio background \citep{waxman00}, and would be
expected to be spatially extended. 
However, it has been found that the fractional fluctuations expected on 
degree scales are of order unity \citep{waxman00}, more than 10 times 
higher than the degree scale fluctuations in
the models discussed above. To sufficiently suppress power on arcminute scales would
therefore require sources even larger than a few Mpc.  As discussed in 
\citet{singal10}, a substantial contribution from diffuse sources may also be limited
by the X-ray and $\gamma$-ray background. 

The calculations in Figure 1 assumed $b=1$. At $z=0$, a galaxy with $b=1$ would have 
a mass near $3 \times 10^{12} h^{-1}M_\odot$, while at $z=5$ this mass corresponds to
a mass below $10^6 M_\odot$. If the typical contribution is coming from higher masses,
the bias factor will be larger, increasing the amplitude of these curves, while to
decrease them by the maximum possible factor of $\sim 0.5$ would require much smaller
masses. Even so, reducing the theoretical curves by a factor of two (requiring all the
CRB to be generated by dwarf galaxies if it is coming from low $z$) is 
still in tension with the ATCA and VLA limits on anisotropy.

\begin{figure}
%\plotone{radio_limits_excess.pdf}
\includegraphics[width=3.5in]{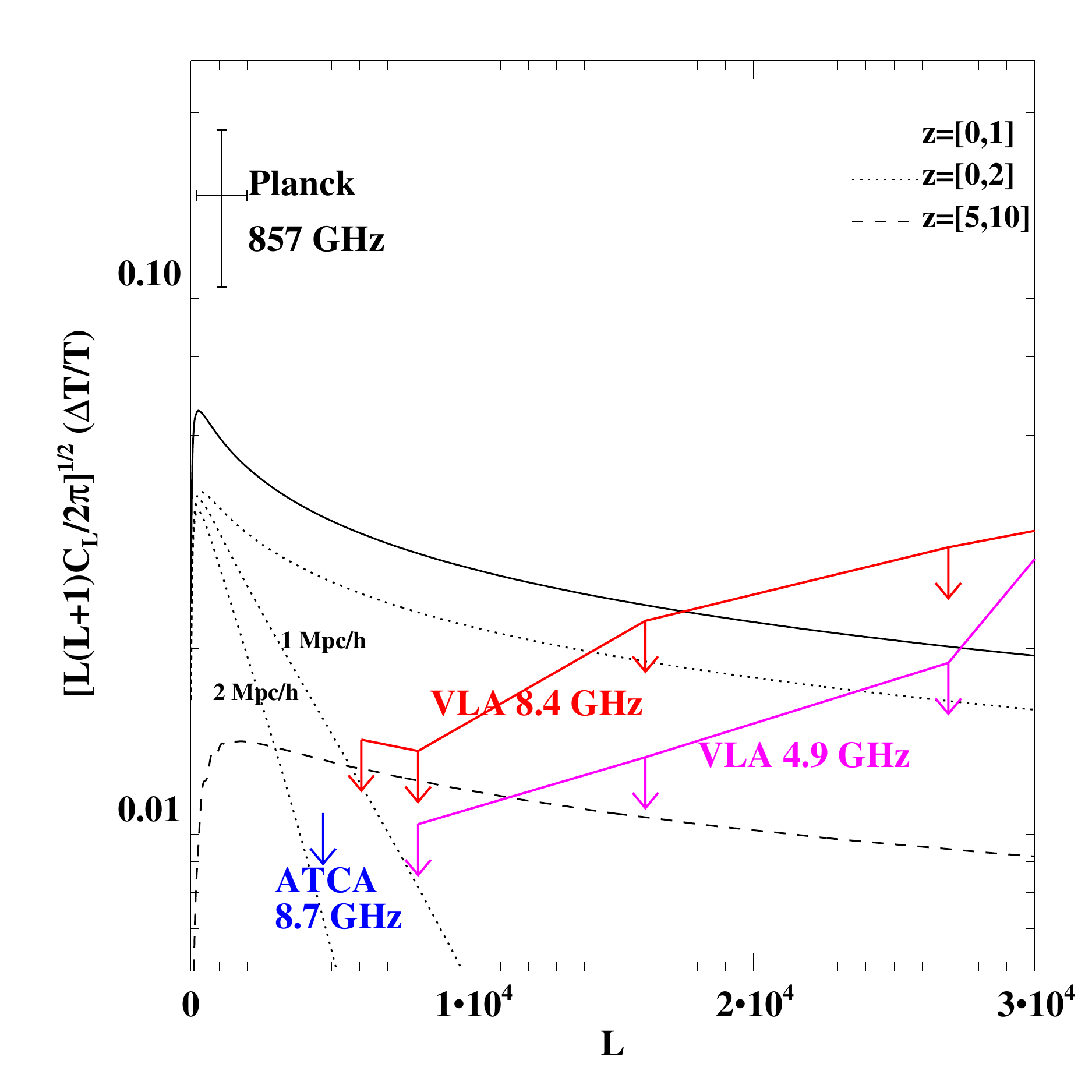}
\caption{Expected clustering for several ranges in redshift for the contributions
to the unresolved radio background, as well as the observed upper limits on clustering
(using the background temperature $T_{excess})$.  
For reference, the amplitude inferred for the cosmic
infrared background measured by Planck is also shown. For the redshift interval
$z=0-2$ (dotted), the effect of each source being extended is shown: 
top to bottom are FWHM$_{smooth}$=0, 1,2,$h^{-1}$comoving Mpc }
\end{figure}

\section{Discussion}

The extragalactic radio background measured by the ARCADE-2 experiment \citep{fixsen11} 
is remarkably smooth, at a level that makes it
unlikely to be generated by emission from a normal population of galaxies. If the
sources are cosmological it would be expected that they trace the large scale structure 
of the universe to some extent. Clustering would therefore generically be
expected to be at the level of a few percent for these sources, in conflict with
upper limits from deep anisotropy searches on small scales. For comparison, the
extragalactic radio background has more than an order of magnitude
smaller rms fluctuations than the cosmic infrared background.

It appears that cosmological sources for this background must be
either at high redshift ($z \ga 5$), where the clustering amplitude is substantially
lower than it is today, or the individual sources must be spatially extended 
(few Mpc in extent), such that there
is not much clustering power on the arcminute scales that have been probed by experiments.

The constraining power of angular clustering measurements is apparent. The upper limits
used in this work are all at least a decade old, with the most stringent fluctuation
limit being almost 25 years old \citep{fomalont88}. With a new generation of radio
experiments, it should be possible to greatly improve on these limits. 
If the clustering amplitude is just below the ATCA/VLA upper limits, this will be
easily measured by an experiment such as the Low Frequency Array (LOFAR). At the upper
end of their frequency range (240 MHz), the excess temperature would be $\sim 40$K. If
this is clustered at even the 0.5\% level, this would produce rms noise fluctuations
$\sim 0.2$K. The typical noise expected for a 1hr observation in the ``Core'' 
configuration is
expected to be 0.5mJy in a $\sim$2' synthesized beam \footnote{www.lofar.org}, 
corresponding to a temperature noise per pixel of roughly 0.8K. The clustering would
thus be a non-negligible fraction of the noise budget for a typical observation, 
and could be easily detected in a dedicated power spectrum measurement with just a
few hours of observation.

If the excess is not caused by a cosmological population, a possibility is
that it is coming from our Galaxy. The radio Galaxy has structure on very large
scales; at 2.3 GHz, the power spectrum of the Galaxy on large scales has been
found to be roughly $C_\ell \sim 0.09$\ K$^2 \ell^{-2.9}$ \citep{giardino01} for
$\ell \lesssim 100$. Extrapolating
this to $\ell \sim 4000$ yields a typical fractional rms temperature fluctuation that is
2\% of the excess temperature, again well above the ATCA and VLA limits. Therefore,
even if the ARCADE-2 measurements are contaminated by the Galaxy this excess is
surprisingly smooth on arcminute scales, requiring the angular power spectrum to
be much steeper than $l^{-3}$ on smaller scales. 

In summary, the high extragalactic temperature measured by the ARCADE-2 experiment
presents a genuine puzzle. Not only is the mean temperature higher than expected based
on extrapolations of source counts, but the small-scale fluctuations in this background 
are much smaller than expected, an order of magnitude smaller than the
fluctuations in the cosmic infrared background. 
Measurements of these fluctuations will be
extremely useful for characterizing the source of this background, and the new generation of
radio experiments is well-equipped to shed new light on this puzzle.

\acknowledgements{This work benefited from the hospitality of
the Kavli Institute for Cosmological Physics, the Fermilab Center for
Particle Astrophysics, and the Aspen Center for Physics. Funding support
came from the Canadian Institute for Advanced Research, the Canada Research
Chairs program, and the NSERC Discovery program. This work benefitted greatly
from constructive discussions with Bud Fisher and Olivier Dore.}

\bibliography{arcade_clustering}
\end{document}